\documentclass[twocolumn,preprintnumbers,amsmath,amssymb,prx]{revtex4}
\usepackage[dvipdfmx]{graphicx}

\usepackage{graphicx}
\usepackage{dcolumn}
\usepackage{bm}
\usepackage{color}
\usepackage{rotating}
\usepackage{ulem} 

\def\SIR{Sr$_2$(Ir$_{1-x}$Rh$_x$)O$_{4}$}

\begin{document}

\title{Bond directional anapole order in a spin-orbit coupled Mott insulator Sr$_2$(Ir$_{1-x}$Rh$_x$)O$_{4}$}

\author{H.\,Murayama$^1$}
\author{K.\,Ishida$^2$}
\author{R.\,Kurihara$^1$}
\author{T.\,Ono$^1$}
\author{Y.\,Sato$^1$}
\author{Y.\,Kasahara$^1$}
\author{H. Watanabe$^1$}
\author{Y. Yanase$^{1,4}$}
\author{G.\,Cao$^3$} 
\author{Y. Mizukami$^2$}
\author{T.\,Shibauchi$^2$}
\author{Y.\,Matsuda$^1$}
\author{S.\,Kasahara$^1$}

\affiliation{$^1$ Department of Physics, Kyoto University, Kyoto 606-8502 Japan}
\affiliation{$^2$ Department of Advanced Materials Science, University of Tokyo, Chiba 277-8561, Japan} 
\affiliation{$^3$  Department of Physics, University of Colorado at Boulder, Boulder, CO 80309, USA}
\affiliation{$^4$ Institute for Molecular Science, Okazaki 444-8585, Japan}



\begin{abstract}
{An anapole state that breaks inversion and time reversal symmetries with preserving translation symmetry of underlying lattice has aroused great interest as a new quantum state, but only a few candidate materials have been reported. Recently, in a spin-orbit coupled Mott insulator \SIR, the emergence of a possible hidden order phase with broken inversion symmetry has been suggested at $T_{\Omega}$ above the N\'{e}el temperature by optical second harmonic generation measurements. Moreover, polarized neutron diffraction measurements revealed the broken time reversal symmetry below $T_{\Omega}$, which was supported by subsequent muon spin relaxation experiments. However, the nature of this mysterious phase remains largely elusive. Here, we investigate the hidden order phase through the combined measurements of the in-plane magnetic anisotropy with exceptionally high-precision magnetic torque and the nematic susceptibility with elastoresistance. A distinct two-fold in-plane magnetic anisotropy along the [110] Ir-O-Ir bond direction sets in below $\sim T_{\Omega}$, providing thermodynamic evidence for a nematic phase transition with broken $C_4$ rotational symmetry. However, in contrast to even-parity nematic transition reported in other correlated electron systems, the nematic susceptibility exhibits no divergent behavior towards $T_{\Omega}$. These results provide bulk evidence for an odd-parity order parameter with broken rotational symmetry in the hidden order state. We discuss the hidden order in terms of an anapole state, in which polar toroidal moment is induced by two current loops in each IrO$_6$ octahedron of opposite chirality. 	Contrary to the simplest loop-current pattern previously suggested, the present results are consistent with a pattern in which the intra-unit cell loop-current flows along only one of the diagonal directions in the IrO$_4$ square. 
}
\end{abstract}
\maketitle

\section{Introduction}
\begin{figure}[t!]
	\centering
	\includegraphics[width=\linewidth]{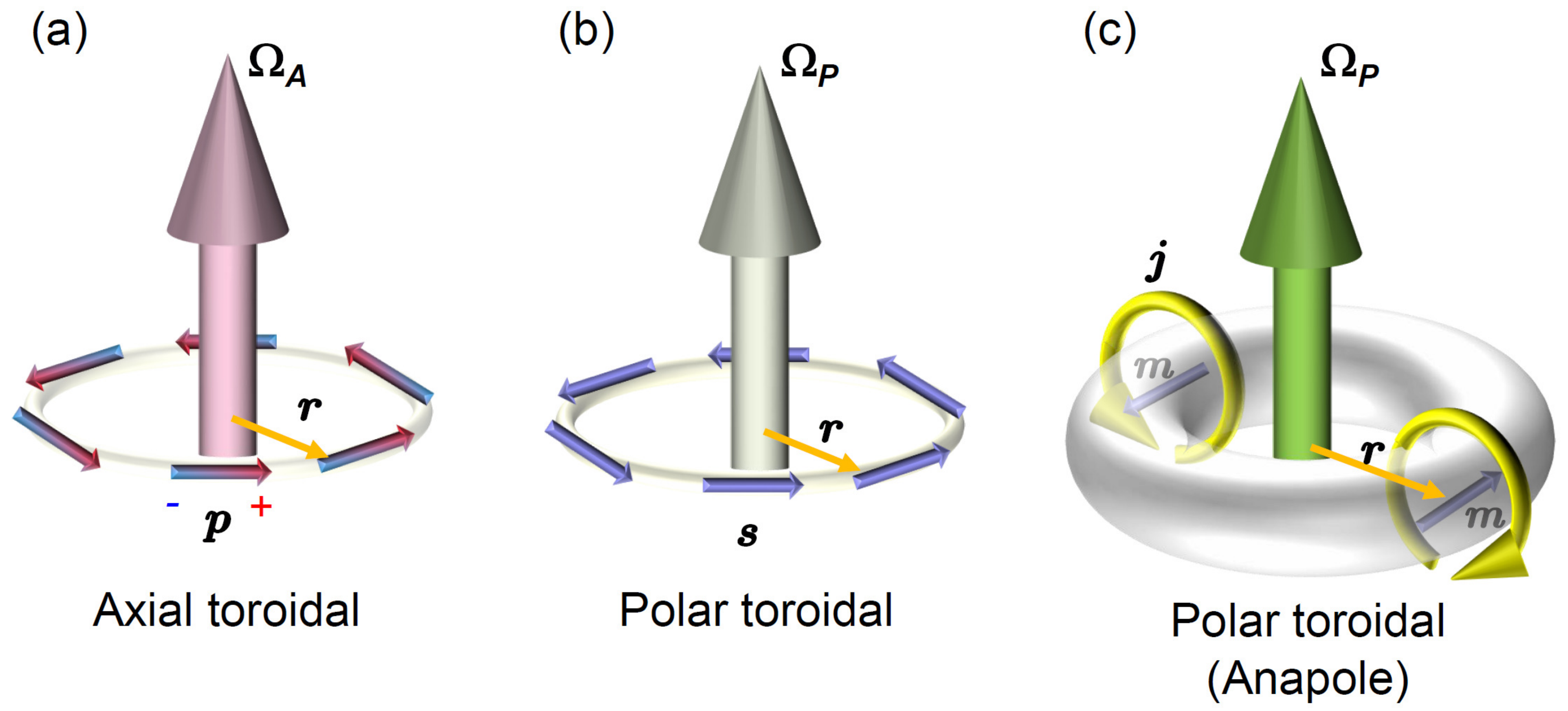}
	\caption{Toroidal states with broken inversion symmetry.   (a) Electric-toroidal state with a vortex-like arrangement of electric dipoles ${\bm p}$.  This is the axial toroidal state in which time-reversal symmetry is preserved. ${\bm \Omega_A}\propto \sum_{i}{\bm r_i} \times {\bm p_i}$ is the axial toroidal moment.  (b) Magneto-toroidal state with a vortex-like arrangement of spin ${\bm s}$.  This is a polar toroidal state in which  time reversal symmetry is broken.  ${\bm \Omega_P}\propto\sum_{i}{\bm r_i} \times {\bm s_i}$ is the  polar toroidal moment.  (c) Anapole state.   
The loop currents ${\bm j}$ flowing the surface of a torus induce toroidal magnetic fields ${\bm m}$.  ${\bm \Omega_P}\propto \int {\bm r} \times {\bm m}({\bm r}) d^3r$ is the anapole vector.
}
	\label{fig:toroidal}
\end{figure}

The search for novel types of ordered state is one of the most exciting and challenging issues of modern condensed-matter physics. In strongly correlated electron systems, charge, spin, and orbital degrees of freedom generate a rich variety of ordered states.   Any of these ordered states can be characterized by their behaviors under space inversion (parity) and time reversal operations.   Among systems with broken parity, toroidal states, in which translational symmetry of underlying lattice is preserved, have been widely discussed~\cite{Spaldin08, Nanz16}. 
There are two types of toroidal states, axial and polar, where time reversal symmetry is preserved and broken, respectively~\cite{Dubovik86,Dubovik90,Gnewuch19}.   An example of the axial toroidal states is the electric-toroidal systems with a vortex-like arrangement of electric dipole~\cite{Yadav16, Jin20, Nakano18}.  The order parameter of this state is represented by electric toroidal dipole moment ${\bm \Omega_A}$,  as  illustrated in Fig.\,\ref{fig:toroidal}(a).   The polar toroidal state is realized in magneto-toroidal systems with a vortex like arrangement of spin and loop currents, as illustrated in Figs.\,\ref{fig:toroidal}(b) and (c), respectively.  The order parameter of these polar toroidal systems is represented by polar magnetic toroidal dipole moment ${\bm \Omega_P}$.   The emergence of polar toroidal dipole moment has aroused significant interest~\cite{VanAken07, Scagnoli11,Matteo12}. In particular, the state, in which polar toroidal dipole is induced by persistent loop currents as illustrated in Fig.\,\ref{fig:toroidal}(c),  is a new quantum state of matter and its finding has been a longstanding quest in condensed matter physics.   To distinguish two polar states shown in Figs.\,\ref{fig:toroidal}(b) and (c),   we call polar toroidal state caused by the loop currents as an anapole state.  The order parameter of the anapole state is represented by anapole vector ${\bm \Omega_P}\propto \int{\bm r} \times {\bm m}({\bm r}) d^3r$, where ${\bm r}$ and  ${\bm m}$ are the position and magnetic moment induced by orbital loop currents, respectively.  For ferrotoroidal coupling in a tetragonal system, where ${\bm \Omega_A}$ or ${\bm \Omega_P}$ is aligned parallel to the $ab$ plane, transnational symmetry is not broken while spatial inversion and four-fold rotational symmetries are broken.  The anapole state has been discussed in the pseudogap state in cuprates~\cite{Varma06,Varma09,Varma14,Pershoguba13-14, Chatterjee17b, 
Scheurer18, Sachdev19,Fauque06,Li08,Almeida-Didry12,
Mangin-Thro17}, but the presence of such a state has been highly controversial~\cite{Croft17, Bourges18, Croft18}.
  
\begin{figure}[t]
	\centering
	\includegraphics[width=\linewidth]{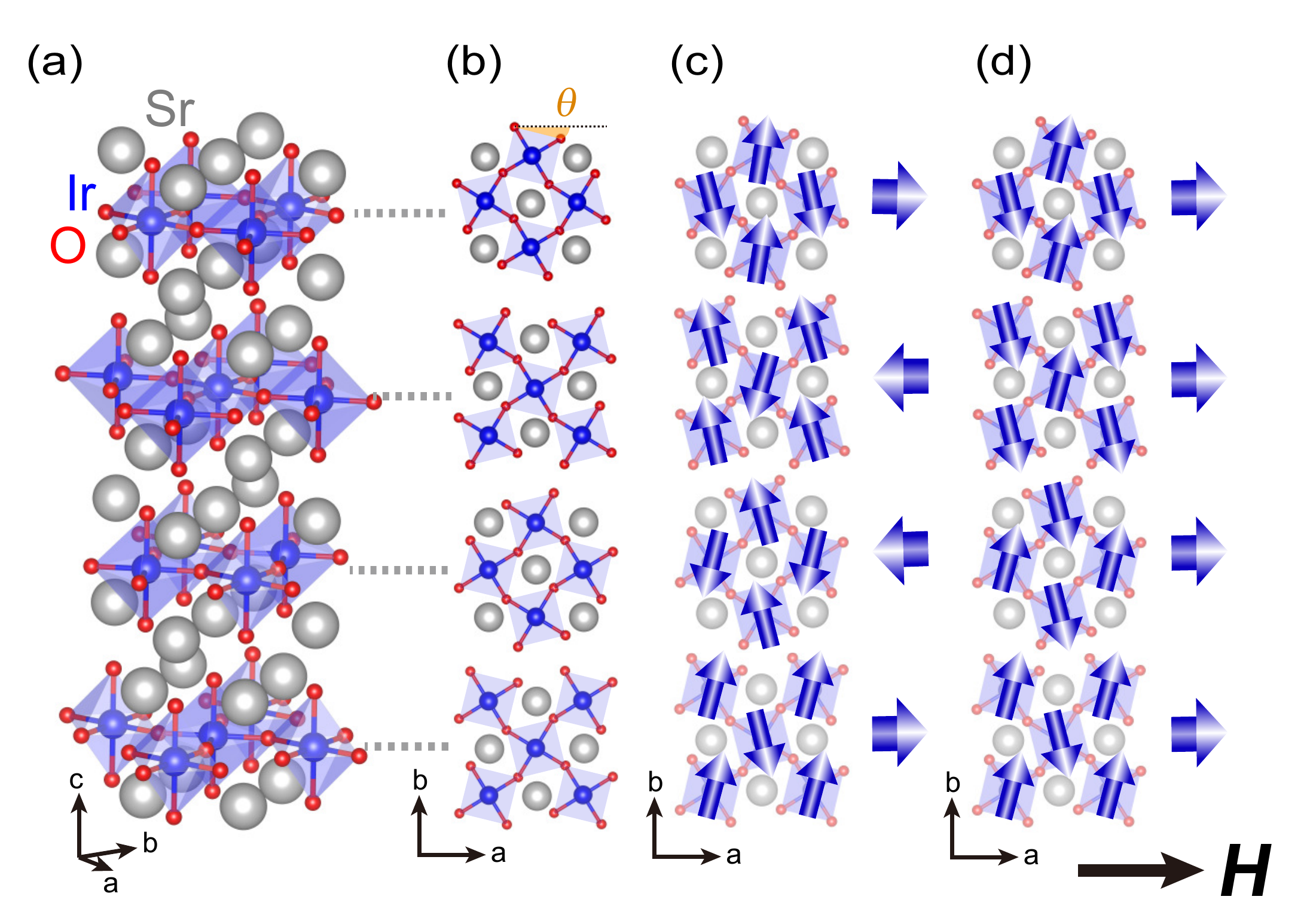}
	\caption{(a) Crystal structure of Sr$_2$IrO$_4$.  (b) Top view of each layer. The crystal has a tetragonal structure with  corner sharing IrO$_6$ octahedra, which is rotated by $\theta=11^{\circ}$ about the $c$ axis. (c)AF-I ($+--+$) magnetic structure. (d) AF-II ($++++$) magnetic structure. Weak ferromagnetic moments are induced in the plane. }
	\label{fig:crystal}
\end{figure}
Recently, the layered perovskite Sr$_2$IrO$_4$ has attracted much interest because it is the first experimental realization of a spin-orbit coupled Mott insulator~\cite{Kim08,Rau16, Cao18, Bertinshaw19,Dai14}. The combination of strong spin-orbit coupling and electron correlation makes this material a promising candidate that hosts a novel electronic state of matter.  The crystal has a tetragonal structure of corner sharing IrO$_6$ octahedra, which is rotated by $\theta=11^{\circ}$ about the $c$ axis [Figs.\,\ref{fig:crystal}(a) and (b)]~\cite{Crawford94, Dhital13, Ye13, Torchinsky15}. A crystalline electric field splits the energy levels of 5$d^5$ electrons in the Ir$^{4+}$ ion into $e_g$ and $t_{2g}$ states.  The presence of strong spin-orbit interaction splits $t_{2g}$ states into a  pseudospin $J_{\rm eff}=1/2$ doublet and  $J_{\rm eff}=3/2$ quartet.  A large enough Coulomb interaction splits the $J_{\rm eff}=1/2$ doublet, leading to a Mott insulating state with one localized electron, which is well described by  $J_{\rm eff}=1/2$ pseudospin anisotropic Heisenberg model~\cite{Kim08}. Such a state has been experimentally established by angle-resolved -photoemission-spectroscopy (ARPES) and 
resonant x-ray scattering experiments~\cite{Kim08, Kim09}.  Below $T_N\approx240$\,K, Sr$_2$IrO$_4$ exhibits a long range antiferromagnetic (AFM) order~\cite{Dhital13,Ye13, Kim09}.  The magnetic moments are aligned in the basal $ab$ plane and  follow the rotation of IrO$_6$ octahedra forming a canted antiferromagnetism with ($+--+$) structure along the $c$ axis (AF-I), as illustrated in Fig.\,2(c)~\cite{Kim09, Boseggia13}.  When a magnetic field of $\gtrsim 0.1$\,T is applied parallel to the $ab$ plane,  the magnetic structure of AF-I changes to that of AF-II with ($++++$) structure shown in Fig.\,\ref{fig:crystal}(d), in which weak ferromagnetic moments are induced within the plane~\cite{Kim09}.   

Despite the complication of the strong spin-orbit interaction, different active orbitals, and structural distortions, it has been suggested that the low-energy electronic structure of Sr$_2$IrO$_4$ bears certain resemblance to that of the cuprates~\cite{Kim08,JKim12}. It has been pointed out that electron-doped iridates should be compared to hole-doped cuprates because of the opposite band curvature due to the opposite sign of the next nearest hopping term~\cite{Wang11}. In electron-doped Sr$_2$IrO$_4$, which is achieved by K-doping at the surface or by partial substitution of La for Sr in the bulk~\cite{Ge11, Chen15}, ARPES and scanning tunneling microscopy measurements report the emergence of Fermi surface pockets, Fermi arcs, and $d$-wave like pseudogap~\cite{Kim14, Kim16, Brouet15, Torre15, Yan15, Battiski17}, but no direct signature of the superconductivity has been reported despite the theoretical predictions~\cite{Wang11,Watanabe13,Yang14,Meng14}.  

\begin{figure}[t]
	\centering
	\includegraphics[width=0.9\linewidth]{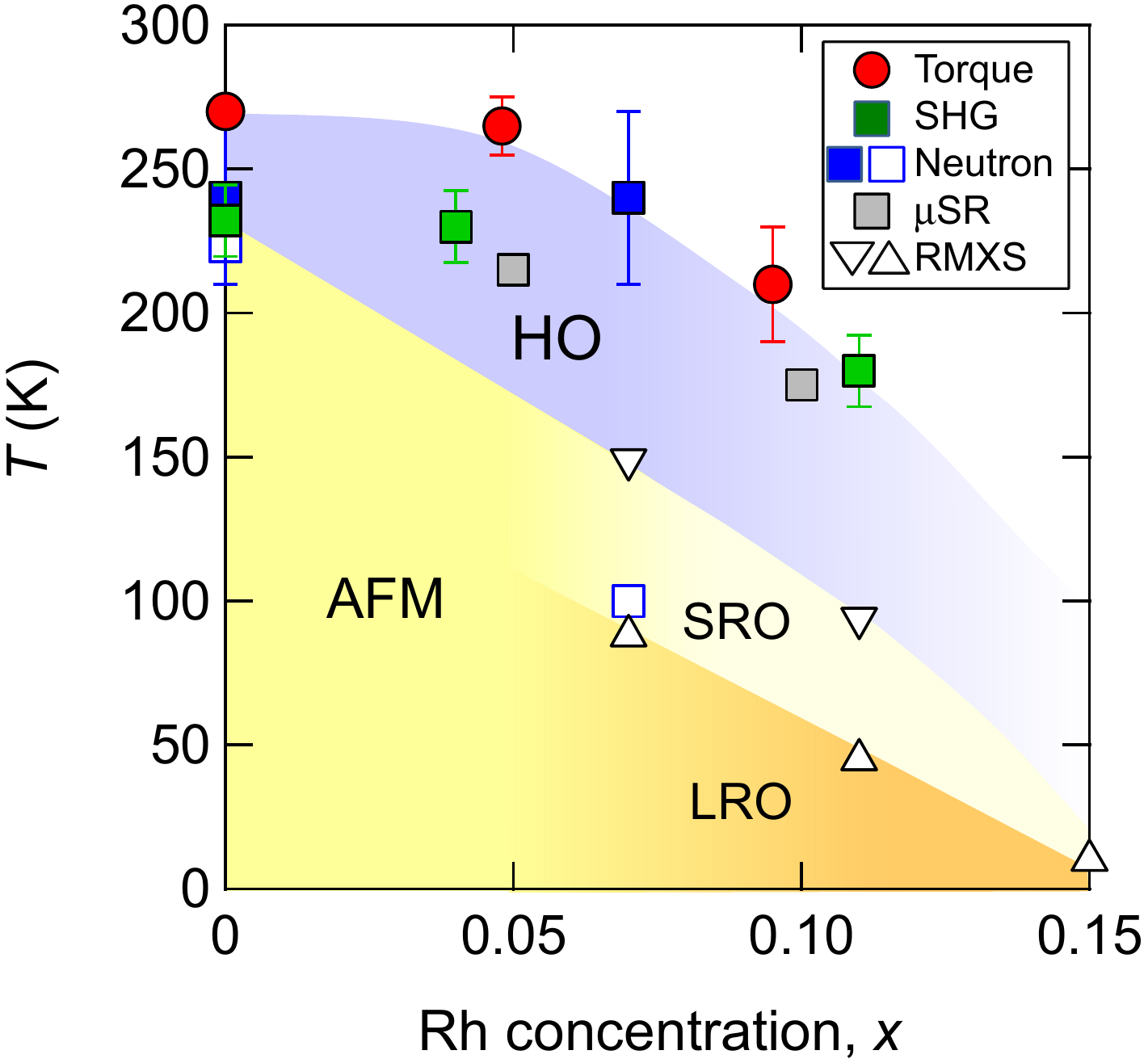}
	\caption{$T$--$x$ phase diagram of \SIR~ determined by the SHG, polarized neutron, resonant magnetic x-ray scattering (RMXS) muon spin relaxation and the present torque measurements. Blue region represents the hidden order (HO) phase.  In the antiferromagnetic (AFM) phase, AF-I magnetic structure at low field changes to AF-II structure in parallel field above $\sim$0.1\,T.  At $x\agt 0.06$, the magnetic short range order (SRO) occurs at high temperature, which is followed by magnetic long range order (LRO) with AF-II structure occurs.}
	\label{fig:phasediagram}
\end{figure}

The effective hole doping is achieved by Rh-substitution~\cite{Qi12,Ye15}.  
It has been shown that Rh ion acts as an acceptor by forming Rh$^{3+}$($4d^6$) oxidation state, creating nearby Ir$^{5+}$(5$d^4$) ions for preserving the charge neutrality~\cite{Clancy14}. The $T$--$x$ phase diagram of \SIR~ is displayed in Fig.\,\ref{fig:phasediagram}. The AFM transition temperature decreases almost linearly as a function of $x$ and vanishes at a critical doping of $x_c\sim 0.17$~\cite{Clancy14}.  
AF-I type magnetic structure is preserved in low $x$ regime.  At $x \agt 0.04$--0.06,  a long-range-ordered (LRO) phase with AF-I type structure is realized  below $T_N^L$, while a short-range-ordered (SRO) phase appears between $T_N^S$ and $T_N^L$~\cite{Clancy14}.  The ARPES measurements report that the $J_{\rm eff}=1/2$ band reaches the Fermi level,  forming small hole pockets around X-point at $x\sim 0.07$--0.10~\cite{Cao16, Louat18}. The pseudogap-like behavior is also reported in the hole-doped Sr$_2$IrO$_4$ by ARPES~\cite{Cao16, Louat18}, although there are several distinct differences in the low energy electronic structure between \SIR~ and  electron-doped cuprates.

\begin{figure}[t]
	\centering
	\includegraphics[width=\linewidth]{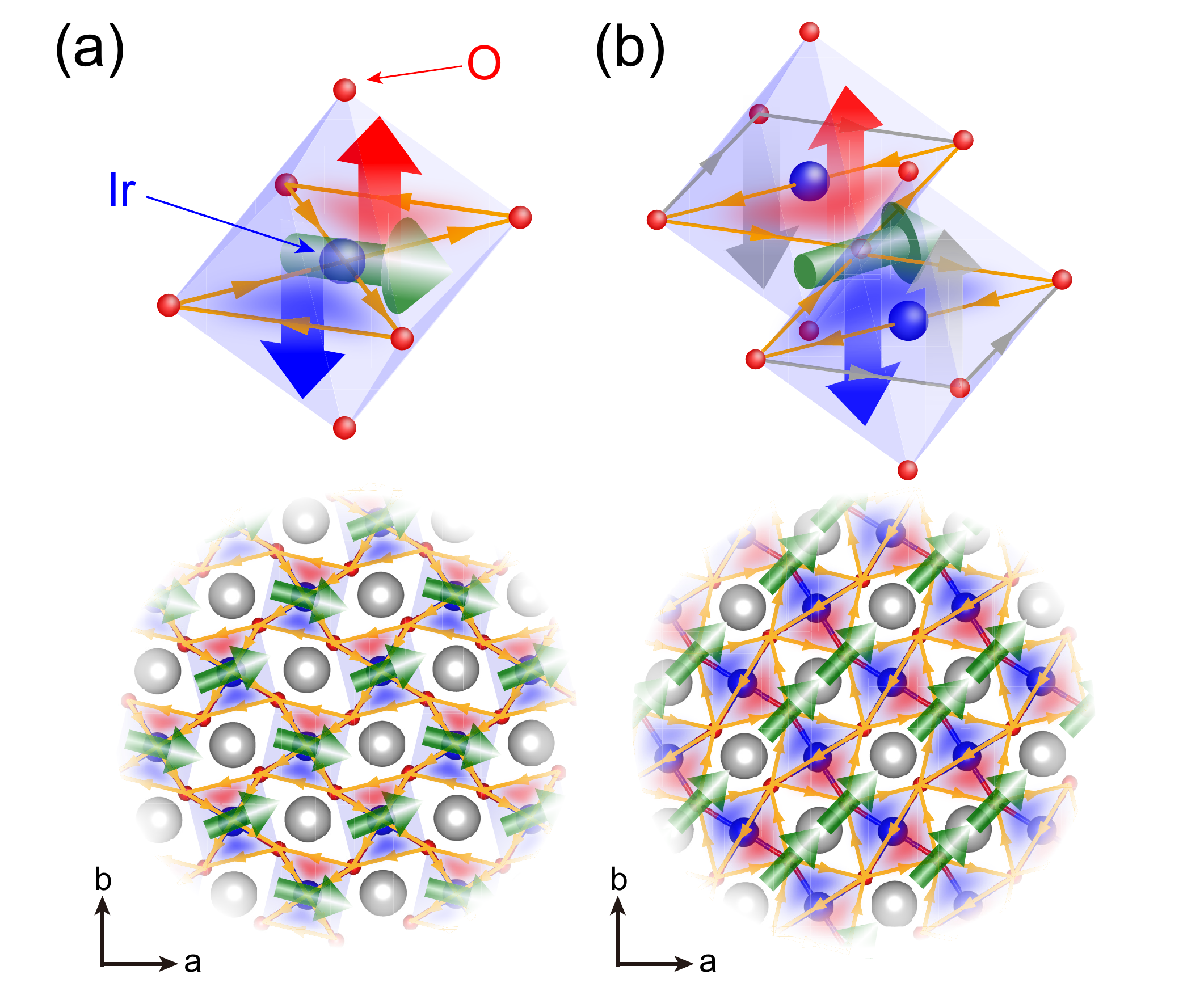}
	\caption{(a) The upper figure illustrates the loop currents (thin orange arrows), induced magnetic fields (thick red and blue arrows) and anapole vector (green arrow) in the IrO$_6$ octahedron. The net anapole vector is along [100] Ir-Ir direction.  The lower figure is the top view of the upper figure.   (b) Same figure as (a), but the loop current pattern is different.  The anapole vector is along the [110] Ir-O-Ir bond direction. }
	\label{fig:anapole}
\end{figure}
The phase diagram of cuprates features a plethora of the broken symmetries in the pseudogap regime that includes charge-density-wave, stripe and nematic and possible loop current orders~\cite{Keimer15}. Therefore, it is highly intriguing to investigate the symmetry breaking phase proximate to the spin-orbit coupled Mott insulator Sr$_2$IrO$_4$.  Recently, the presence of a hidden broken symmetry phase that develops prior to the AFM phase has been reported in pure and Rh-substituted Sr$_2$IrO$_4$ by optical second harmonic generation (SHG) measurements~\cite{Zhao16}. The SHG can sensitively detect odd-parity order parameter but is insensitive to even-parity order parameter~\cite{Zhao16,Zhao17,Harter17}. In Fig.\,\ref{fig:phasediagram}, the onset temperature of the hidden order determined by the SHG signal  $T_\Omega$ is plotted.  For pure Sr$_2$IrO$_4$, $T_\Omega$ appears to be a few Kelvin above $T_N$ and for Rh-doped Sr$_2$IrO$_4$, $T_\Omega$ is distinctly far above $T_N$. These results suggest that an odd parity hidden order phase develops at higher temperatures above the AFM transition.  

Subsequent polarized neutron diffraction measurements in pure and 7\% Rh-doped Sr$_2$IrO$_4$ report the development of an unconventional magnetic order that breaks time reversal symmetry above  $\sim T_{N}$, which is  characterized by  an intra-unit-cell magnetic order~\cite{Jeong17}.  The onset temperature of this order matches $T_\Omega$ determined by SHG.   It should be noted that similar unconventional magnetic order has been reported in the pseudogap state of underdoped cuprates, including  YBa$_2$Cu$_3$O$_{6+\delta}$\cite{Fauque06},  HgBa$_2$CuO$_{4+\delta}$\cite{Li08}, and Bi$_2$Sr$_2$CaCu$_2$O$_y$\cite{Almeida-Didry12}. The intra-unit-cell magnetism has been discussed in terms of possible counter-circulating loop currents within the unit cell of IrO$_2$ or CuO$_2$ planes (Varma's loop)~\cite{Varma06,Varma09,Varma14,Pershoguba13-14}.   Very recent muon spin relaxation ($\mu$SR) measurements on \SIR~ with $x = 0.05$ and 0.1 report the critical slowing down of electronic spin fluctuations at $\sim T_{\Omega}$~\cite{Tan20}.   Similar phenomena have been reported in the pseudogap state of cuprates~\cite{LeiShu}.  It has been claimed that these muon results are consistent with the polarized neutron experiments.  The onset temperatures of the unconventional magnetic order reported by neutron and $\mu$SR measurements are plotted in Fig.\,\ref{fig:phasediagram}.  Since there is no evidence of structural distortions that breaks translational symmetry above $T_N$, the hidden order state is suggested to be a possible anapole state.

Despite the SHG, polarized neutron diffraction, and $\mu$SR measurements, however, the nature of the hidden order state remains largely elusive.  In fact, because the SHG is a surface probe, it is crucially important to clarify the spatial symmetry breaking by a thermodynamic bulk probe.  
Moreover, although polarized neutron measurements suggest that the direction of the magnetic moment is perpendicular to the $ab$ plane, the direction of anapole vector is unknown~\cite{Jeong17}. As there are several patterns of the loop current~\cite{Varma06,Varma09,Varma14,Pershoguba13-14,Chatterjee17b, 
Scheurer18, Sachdev19}, determining the direction of anapole vector is essentially important for understanding the origin of the broken time reversal symmetry.  
In Figs.\,\ref{fig:anapole}(a) and (b), two examples of the anapole vector and loop-current patterns are illustrated.

In this paper, to obtain deep insight into the symmetry breaking phenomena in the hidden order phase of \SIR, we measure the magnetic susceptibility anisotropy within the IrO$_2$ plane with exceptionally high precision magnetic torque experiments~\cite{Okazaki11, Kasahara12, Sato17, Murayama19} and nematic susceptibility with elastoresistance~\cite{Chu12, Kuo16, Hosoi16, Shapiro16, Ishida20a, Ishida20b}.   
We find that a distinct $C_2$ in-plane anisotropy which breaks $C_4$ symmetry of the underlying lattice sets in below $\sim T_{\Omega}$ in \SIR.  This provides thermodynamic evidence for a nematic transition, which bears resemblance to the phenomena observed in the pseudogap state of cuprates~\cite{Sato17, Murayama19}. However, the nematic director of \SIR~ is along the Ir-O-Ir bond direction, in stark contrast to that of monolayer cuprate HgBa$_2$CuO$_{4+\delta}$, which is 45$^\circ$ rotated from the Cu-O-Cu bond direction~\cite{Murayama19}. Moreover, in contrast to  iron-based superconductors, in which the nematic susceptibility is largely enhanced towards  the nematic transition~\cite{Chu12, Kuo16, Hosoi16, Shapiro16, Ishida20a}, the nematic susceptibility of \SIR~exhibits no divergent behavior towards $T_\Omega$, which is consistent with the odd parity order parameter. Based on these results, we discuss the hidden order state in terms of an anapole state with peculiar loop current patterns.

\begin{figure}[t]
	\centering
	\includegraphics[width=0.9\linewidth]{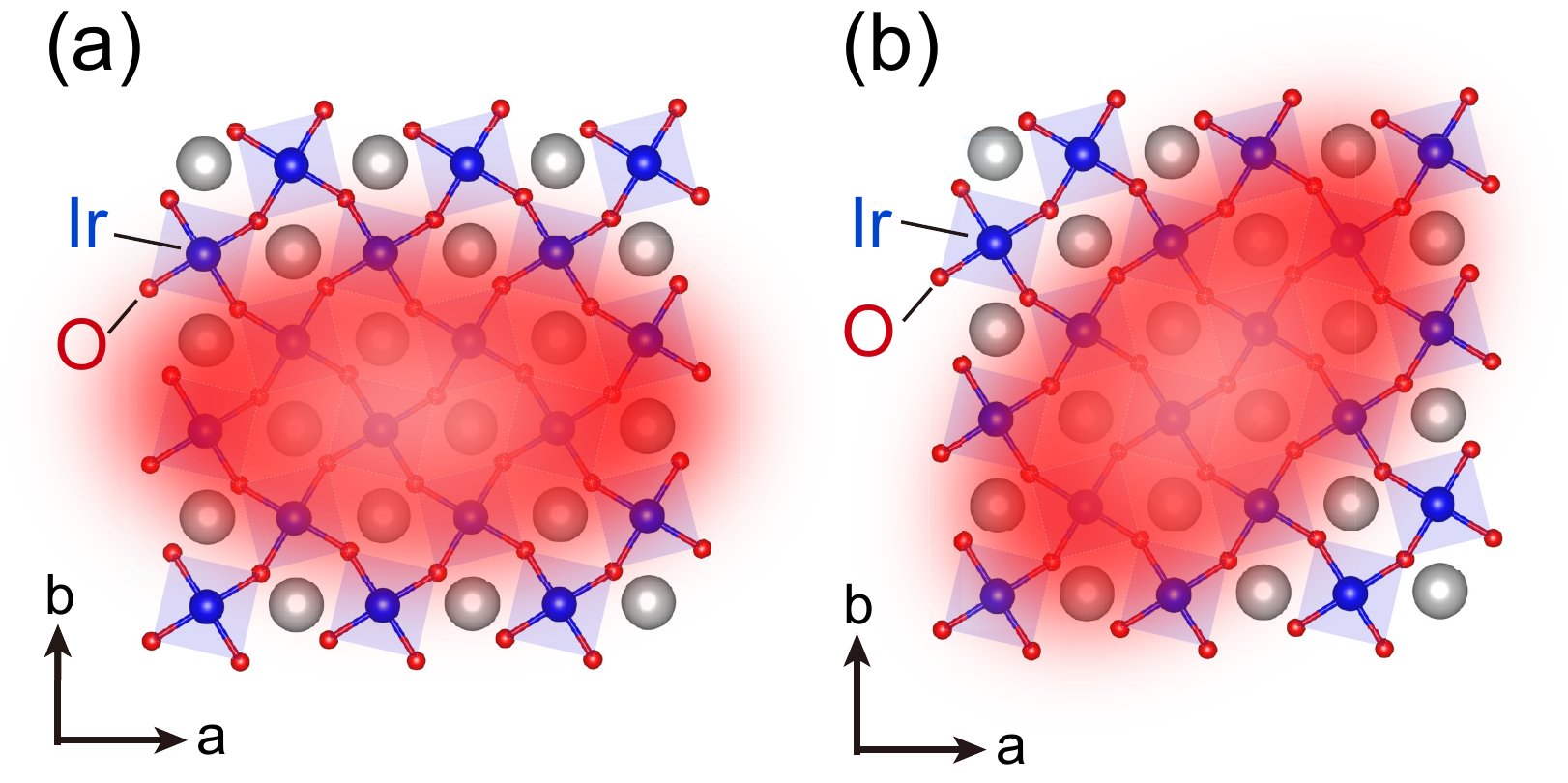}
	\caption{(a) Schematic picture of the nematicity with $B_{1g}$ symmetry, where the nematic director is along the Ir-Ir  [100] direction.  For this nematicity,$\chi_{aa}\neq\chi_{bb}$ and $\chi_{ab}=0$.  (b) The nematicity with $B_{2g}$ symmetry, where the nematic director is along the [110] Ir-O-Ir direction.   For this nematicity, $\chi_{aa}=\chi_{bb}$ and $\chi_{ab}\neq0$}
	\label{fig:nematic}
\end{figure}
\begin{figure}[t]
	\centering
	\includegraphics[width=\linewidth]{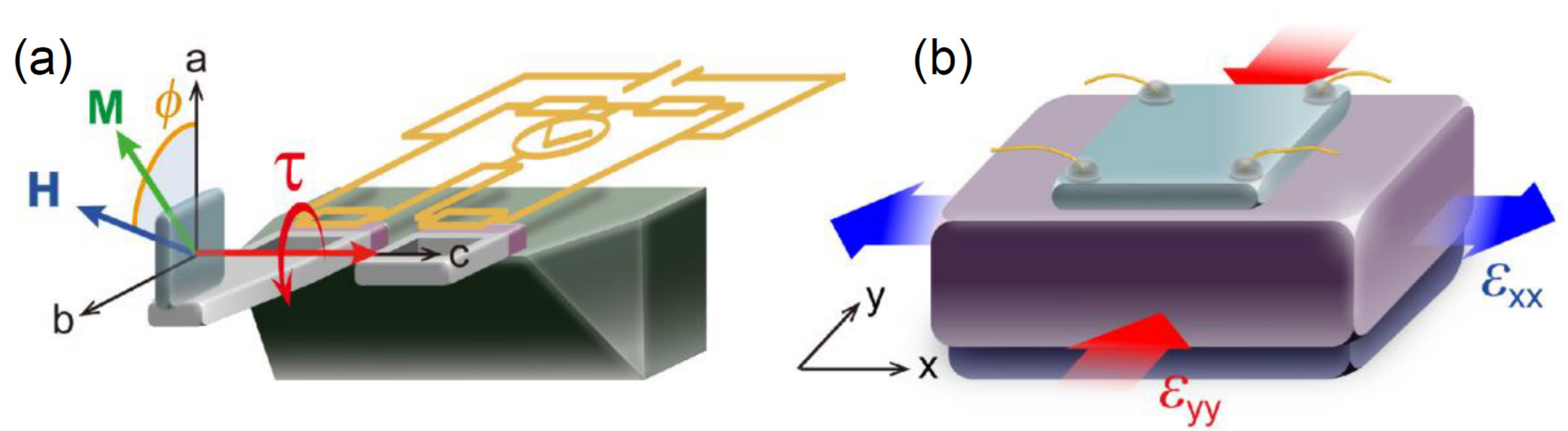}
	\caption{(a) Experimental set-up of in-plane magnetic torque measurements. Magnetic torque ${\bm \tau}$=$\mu_0 {\bm H}\times {\bm M}$ is detected by micro cantilever.  $ {\bm H}$ is rotated within the $ab$ plane. (b) Experimental set-up of elastoresistance measurements. The sample (light blue) is glued on the piezo stack (purple) and the resistances $R_{xx}$ and $R_{yy}$ along $x$ and $y$ directions, respectively, are measured by the van der Pauw method. The strain $\varepsilon_{xx}$ along $x$ is measured by a strain gauge attached on the back side of the piezo stack. The strain $\varepsilon_{yy}$ along $y$ is calculated by the temperature-dependent Poisson ratio, which has been calibrated in advance.}
	\label{fig:experimental}
\end{figure}

\section{Experimental}

Measurements of the magnetic torque ${\bm \tau}=\mu_0V{\bm M}\times {\bm H}$ have a high sensitivity for the detection of magnetic anisotropy, where $\mu_0$ is space permeability, $V$ is the sample volume, and $\bm{M}$ is the magnetization induced by external magnetic field ${\bm H}$. Torque is a thermodynamic quantity that is given by the derivative of the free energy with respect to angular displacement, and can thus shed light on the thermodynamic character of the transition.  We performed the torque measurements  for a range of directions of ${\bm H}$ within the tetragonal $ab$ plane of \SIR~ to test whether the hidden order breaks the four-fold crystal symmetry. In this configuration, ${\bm \tau}$ is a periodic function of twice
the azimuthal angle $\phi$ measured from the $a$ axis:
\begin{equation}
\tau_{2\phi}=\frac{1}{2}\mu_0H^2[(\chi_{aa}-\chi_{bb})\sin 2\phi-2\chi_{ab}\cos 2\phi]
\end{equation}
where $\chi_{ij}$ is the susceptibility tensor defined as $M_{ij}=\sum_j\chi_{ij}H_j$ ($i,j=a,b,c$). For a tetragonally symmetric system,  $\tau_{2\phi}$ should be zero because $\chi_{aa}=\chi_{bb}$ and $\chi_{ab}=0$. Nonzero values of $\tau_{2\phi}$ appear when the tetragonal symmetry is broken by a new electronic or magnetic state; $C_4$ rotational symmetry breaking is revealed by $\chi_{aa}\neq\chi_{bb}$ and/or $\chi_{ab}\neq 0$. The former and the latter states are illustrated in Figs.\,\ref{fig:nematic}(a) and (b), respectively, where the $C_4$ symmetry breaking occurs along [100]/[010] direction ($B_{1g}$-symmetry) and [110] direction (bond directional nematicity with $B_{2g}$-symmetry) of the IrO$_2$ plane.
 \begin{figure*}[t]
	\centering
	\includegraphics[width=\linewidth]{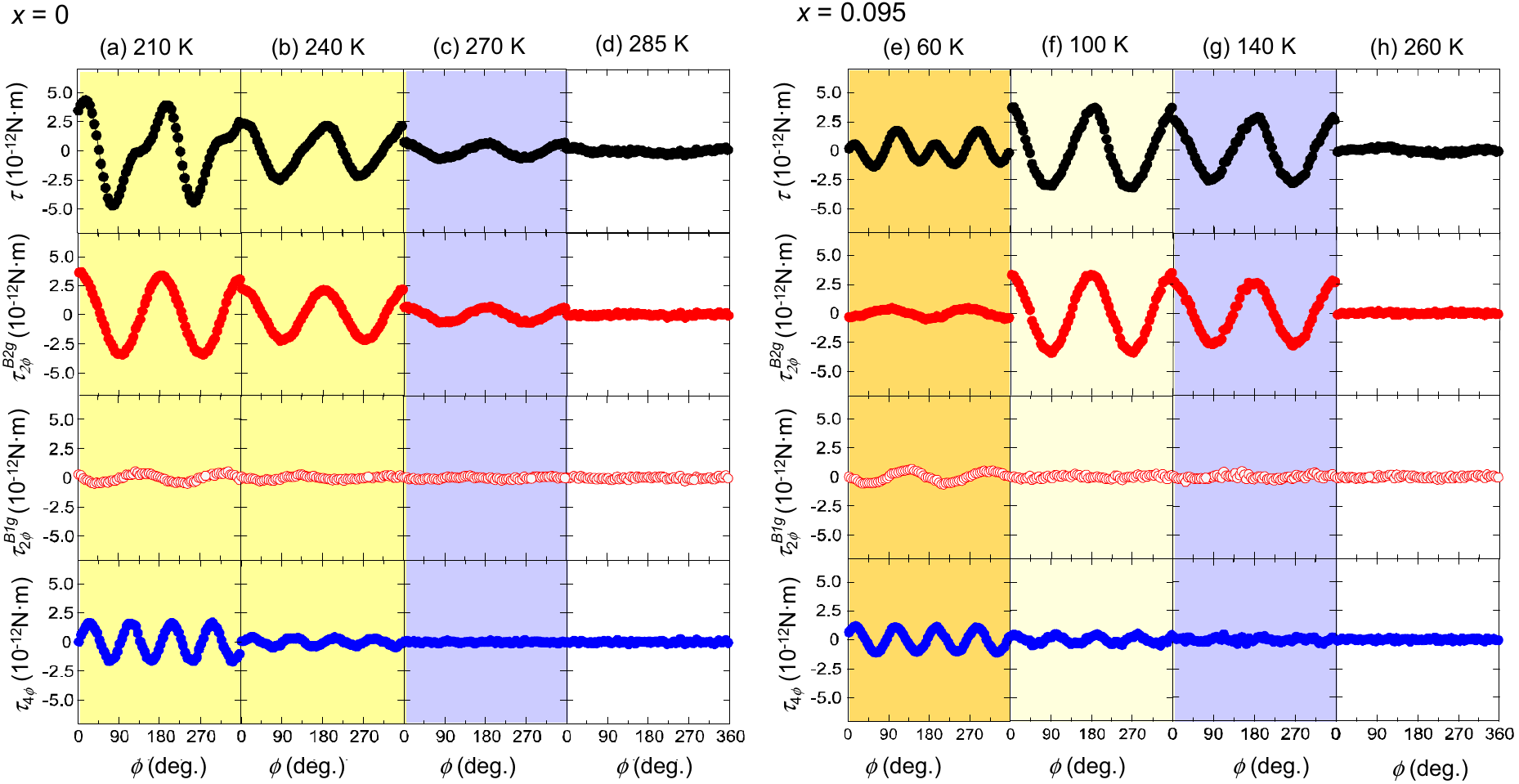}
	\caption{(a)-(d) and (e)-(h) display the  magnetic torque $\tau(\phi)$ in in-plane magnetic field of $|{\mu_0\bm H}|=4$\,T as a function of the azimuthal angle $\phi$ for pure Sr$_2$IrO$_4$ ($T_N \approx 240$\,K) and Sr$_2$Ir$_{0.905}$Rh$_{0.095}$O$_4$ ($T_N^S \approx 120$\,K and $T_N^L \approx 80$\,K), respectively. For $x = 0$, white, blue and yellow shaded regions show that the system is in the paramagnetic, hidden order, and AFM states, respectively. For $x=0.095$, orange (light yellow) region shows that the system is in the long range (short range) AFM state. Top panels show raw torque curves $\tau(\phi)$. Upper middle, lower middle and bottom panels show the $B_{2g}$ two-fold ($\cos2\phi$), $B_{1g}$ two-fold ($\sin2\phi$) and four-fold ($\sin4\phi$) components of the torque curves, respectively, obtained from Fourier analysis of the raw torque curves. 
}
	\label{fig:oscillation}
\end{figure*}

For the in-plane torque magnetometry, we used high sensitive piezoresistive cantilever~\cite{Okazaki11,Kasahara12,Sato17,Murayama19}. The experimental set-up for this measurement is illustrated in Fig.\,\ref{fig:experimental}(a). To measure the in-plane variation of the magnetic torque, we used a system with two superconducting magnets generating magnetic fields in two mutually orthogonal directions and cryostat set on a mechanical rotating stage at the top of a dewar~\cite{Okazaki11,Kasahara12,Sato17,Murayama19}. By computer-controlling the two superconducting magnets and rotating stage, ${\bm H}$ is precisely applied and rotated within the $ab$ plane.  We carefully checked the misalignment of ${\bm H}$ from the $ab$ plane by rotating  ${\bm H}$ conically about the $c$ axis and  found the misalignment is less than 0.1$^{\circ}$.  When the nematicity appears in the tetragonal lattice, the formation of nematic domains is naturally expected.   In the presence of large number of domains, the two-fold oscillations due to nematicity are canceled out.   According to the SHG measurements, the typical domain size is 50--100\,$\mu$m \cite{Zhao16}.  Therefore, we used crystals with size of $\lesssim$100\,$\mu$m  in the torque measurements. Since such crystals contain very small number of domains, the two-fold oscillations can be detected if nematicity appears.

The nematic susceptibility is determined by measuring the elastoresistance. The experimental set-up for this measurement is illustrated in Fig.\,\ref{fig:experimental}(b). Two samples are cut from a single crystal of Sr$_2$Ir$_{0.88}$Rh$_{0.12}$O$_4$ and the crystal axes of the samples are determined by X-ray diffraction. Four contacts are attached near the corners of the samples by silver paste and cured at $\sim250^\circ$C. The two samples are glued on the piezo stack, whose $x$ direction is aligned along [100] and [110] directions, respectively. The van der Pauw method is used to measure resistance changes $\Delta R_{xx}$ and $\Delta R_{yy}$ along $x$ and $y$ direction of the piezo stack~\cite{Shapiro16,Ishida20a}. The difference 
$\eta=(\Delta R/R)_{xx} - (\Delta R/R)_{yy}$ 
is the anisotropy induced by the anisotropic strain $\varepsilon_{xx}-\varepsilon_{yy}$, which is measured by the strain gauge. The quantity $\eta$ is expected to be intimately linked to an electronic nematic order parameter. When the anisotropic strain couples linearly to the nematic order, the nematic susceptibility can be defined as the quantity 
\begin{equation}
\chi_{\rm nem} \equiv d\left[(\Delta R/R)_{xx} - (\Delta R/R)_{yy}\right]/d(\varepsilon_{xx}-\varepsilon_{yy}).
\end{equation}
The measurements of two directions along [100] and [110] correspond to two irreducible representations ($B_{1g}$ and $B_{2g}$) for the tetragonal $D_{4h}$ system.

\section{Results}
\subsection{Magnetic torque}

Top panels of Figs.\,\ref{fig:oscillation}(a)-(d) and (e)-(h) display the raw data of the  magnetic torque $\tau(\phi)$ in magnetic field of $\mu_0$$|{\bm H}|=4$\,T rotating within the $ab$ plane at several temperatures for pure Sr$_2$IrO$_4$ and Sr$_2$Ir$_{0.905}$Rh$_{0.095}$O$_4$ ($T_N^S\approx 120$\,K, $T_N^L\approx 80$\,K), respectively. Note that we use the unit cell with the space group $I4_{1}/a$, where $a$ axis corresponds to Ir-Ir direction, as shown in Figs.\,\ref{fig:crystal}(a) and (b). This is 45$^{\circ}$ rotated from the unit cell of cuprates, where $a$ axis corresponds to Cu-O-Cu bond direction.  All torque curves are perfectly reversible with respect to the field rotation direction, indicating no detectable ferromagnetic impurities.  For both crystals, no oscillations are observed in $\tau(\phi)$ at the highest temperatures, which is consistent with the tetragonal crystal symmetry. At lower temperatures, $\tau(\phi)$ exhibits distinct angular dependence.  We analyze $\tau(\phi)$ by decomposing as $\tau = \tau_{2\phi} + \tau_{4\phi}$, where $\tau_{n\phi} = A_{n\phi}\sin\left[n(\phi-\phi_{n0})\right]$ is a term with $n$-fold symmetry. The two-fold component is further decomposed as $\tau_{2\phi} = A_{2\phi}^{B1g}\sin 2\phi + A_{2\phi}^{B2g}\cos 2\phi$. In the upper middle, lower middle and bottom panels of Figs.\,\ref{fig:oscillation}(a)-(h), the $B_{2g}$ two-fold, $B_{1g}$ two-fold and four-fold components obtained from the Fourier analysis of the raw torque curves are displayed.  The raw $\tau(\phi)$ curves are well reproduced by the sum of these three components. 

\begin{figure}[t]
	\centering
	\includegraphics[width=\linewidth]{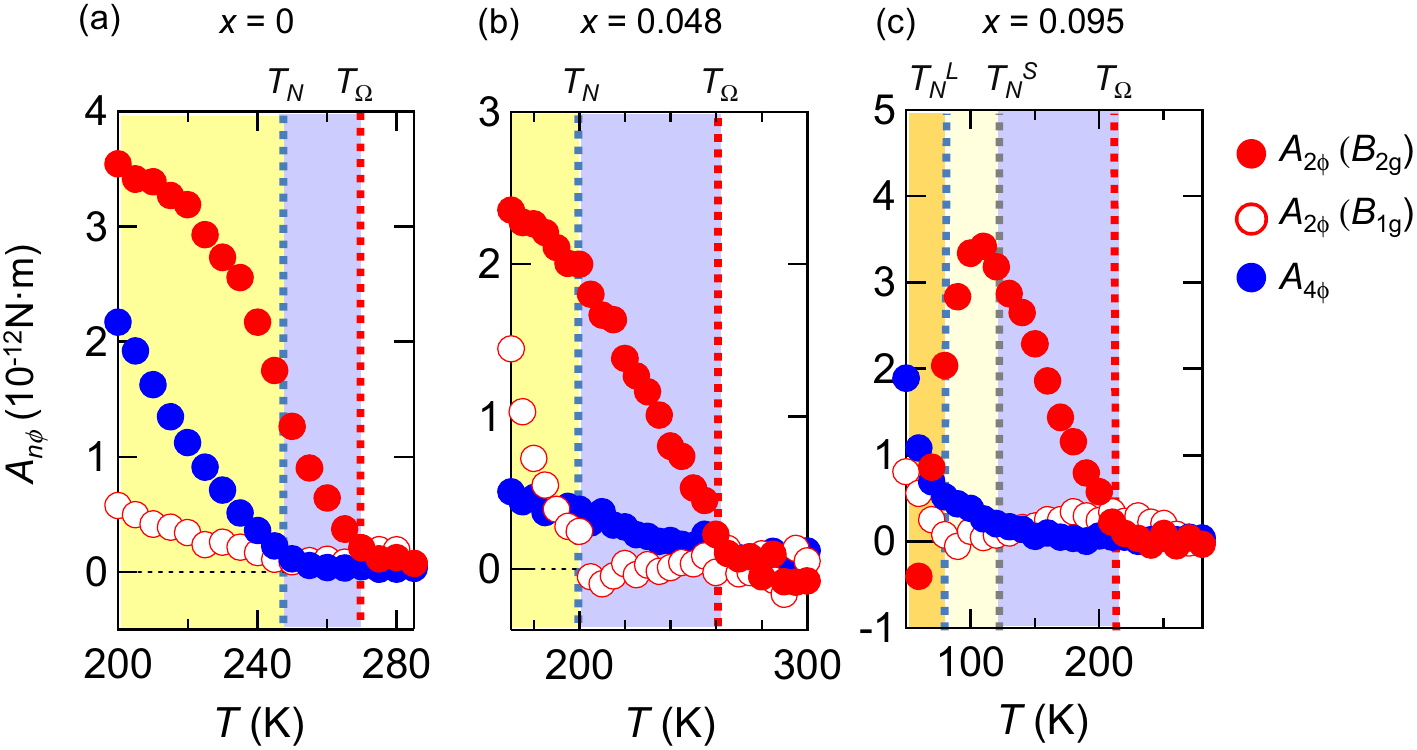}
	\caption{Temperature dependence of the amplitude of two-fold (red filled and open circles)  and  four-fold oscillations (filled blue circles) for pure and Rh-doped Sr$_2$IrO$_4$.  (a) $x=0$, (b) $x=0.048$, (c) $x=0.095$. }
	\label{fig:amplitude}
\end{figure}

As shown in the upper middle panels of Figs.\,\ref{fig:oscillation}(a)-(g) distinct two-fold oscillation with $B_{2g}$ symmetry ($\propto \cos 2\phi$)  is observed  at 270\,K for Sr$_2$IrO$_4$ and at 140\,K for Sr$_2$Ir$_{0.905}$Rh$_{0.095}$O$_4$. It should be stressed that these temperatures are obviously higher than $T_N$. Therefore, the data provide the first evidence for a phase transition to an electronic nematic state as a bulk property. As shown in the lower middle and bottom panels of Figs.\,\ref{fig:oscillation}(a)--(h), small but finite two-fold oscillations with $B_{1g}$ symmetry ($\propto \sin 2\phi$) and four-fold oscillations, which has the form $\tau_{4\phi}=A_{4\phi}\sin 4\phi$, are observed at low temperatures.

 Figures\,\ref{fig:amplitude}(a),(b), and (c) depict the temperature dependence of the amplitude of each oscillation for  \SIR~with $x=0$ (pure), 0.048, and 0.095, respectively.  In all crystals,  finite two-fold $B_{2g}$ oscillations appear well above $T_N$ or $T_N^S$.  Moreover, the amplitude of two-fold oscillations grows nearly linearly with decreasing temperature as $A_{2\phi}^{B2g}\propto |T_{\Omega}-T|^{\beta}$ with $\beta\approx 1$. The onset temperature of the two-fold $B_{2g}$ oscillations are plotted in the $T$--$x$ phase diagram displayed in Fig.\,\ref{fig:phasediagram}.   For pure Sr$_2$IrO$_4$, the onset temperature of two-fold $B_{2g}$ oscillations is slightly higher than the onset temperatures reported by the SHG and polarized neutron scattering measurements~\cite{Zhao16,Jeong17}.  In Rh-doped Sr$_2$IrO$_4$, on the other hand, the onset temperature appears to be close to that determined by the SHG and neutron measurements.  Considering the ambiguity in determining the onset temperature of the SHG signal and the neutron diffraction intensity, the onset temperature of the tetragonal $C_4$ symmetry breaking appears to match that of the hidden order $T_\Omega$. Thus, the present results provide thermodynamic evidence of the nematic transition at $\sim T_\Omega$.  
 
 For pure Sr$_2$IrO$_4$ and $x=0.048$, $A_{2\phi}^{B2g}(T)$ increases with decreasing temperature with no discernible change at $T_N$ and shows a saturating behavior at lower temperatures. These results suggest that the nematicity is insensitive to the long range AFM order. In contrast, for 0.095, $A_{2\phi}^{B2g}(T)$  is strongly suppressed below  $\sim T_N^S $ and  changes the sign below $T_N^L$, suggesting that the $B_{2g}$ nematicity is strongly influenced by the appearance of short-range magnetic structure.   With decreasing temperature, the amplitude of four-fold oscillations increases below $T_N$ for pure Sr$_2$IrO$_4$~\cite{Fruchter12,Nauman17}. The torque curves are perfectly reversible with respect to the field rotation below $T_N$, indicating that the induced weak ferromagnetic moments completely follows the applied ${\bm H}$ rotated with in the $ab$ plane.  
Moreover, the four-fold oscillations for $x = 0.048$ and 0.095 appear even above the AFM ordering temperatures. These results indicate that the four-fold oscillations are not caused by the weak ferromagnetic moments shown in Fig.\,\ref{fig:crystal}(d), but arise primarily from the nonlinear susceptibilities~\cite{Okazaki11}.

\subsection{Elastoresistance}
The elastoresistance measurements have been used as a powerful probe of electronic nematic transitions in strongly correlated electron systems, especially in iron-based superconductors~\cite{Chu12,Kuo16,Hosoi16,Ishida20a, Shibauchi20}. In a condition that the nematic order couples linearly to the strain, this technique can quantify the nematic susceptibility, which is related to the fluctuations of a rotational symmetry breaking order above the nematic transition. Thus, this is complementary to the magnetic torque measurements, which measure the nematic order parameter below the transition. 

  \begin{figure}[t]
	\centering
	\includegraphics[width=\linewidth]{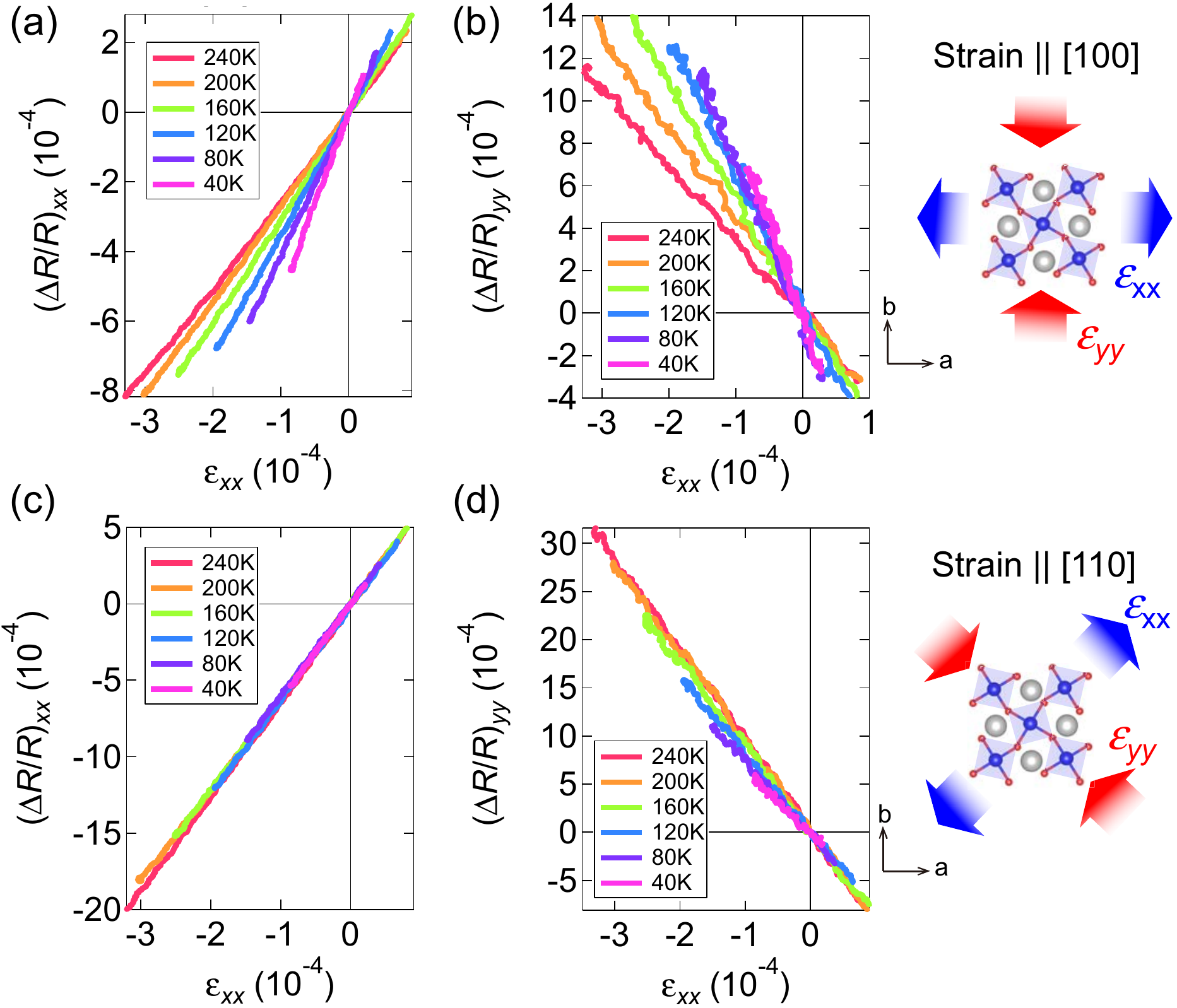}
	\caption{Representative data of elastoresistance for Sr$_2$(Ir$_{1-x}$Rh$_x$)O$_4$ with $x=0.12$. (a,\,b) Longitudinal elastoresistance $(\Delta R/R)_{xx}$ (a) and transverse elastoresistance $(\Delta R/R)_{yy}$ (b) for several temperatures as a function of strain $\varepsilon_{xx}$ with the strain direction $x$ parallel to [100]. (c,\,d) Longitudinal elastoresistance $(\Delta R/R)_{xx}$ (c) and transverse elastoresistance $(\Delta R/R)_{yy}$ (d) for several temperatures as a function of strain $\varepsilon_{xx}$ with the strain direction $x$ parallel to [110]. Linear response of the resistance against the strain can be seen in all the elastoresistance data. }
	\label{fig:strain}
\end{figure}

\begin{figure}[t]
	\centering
	\includegraphics[width=\linewidth]{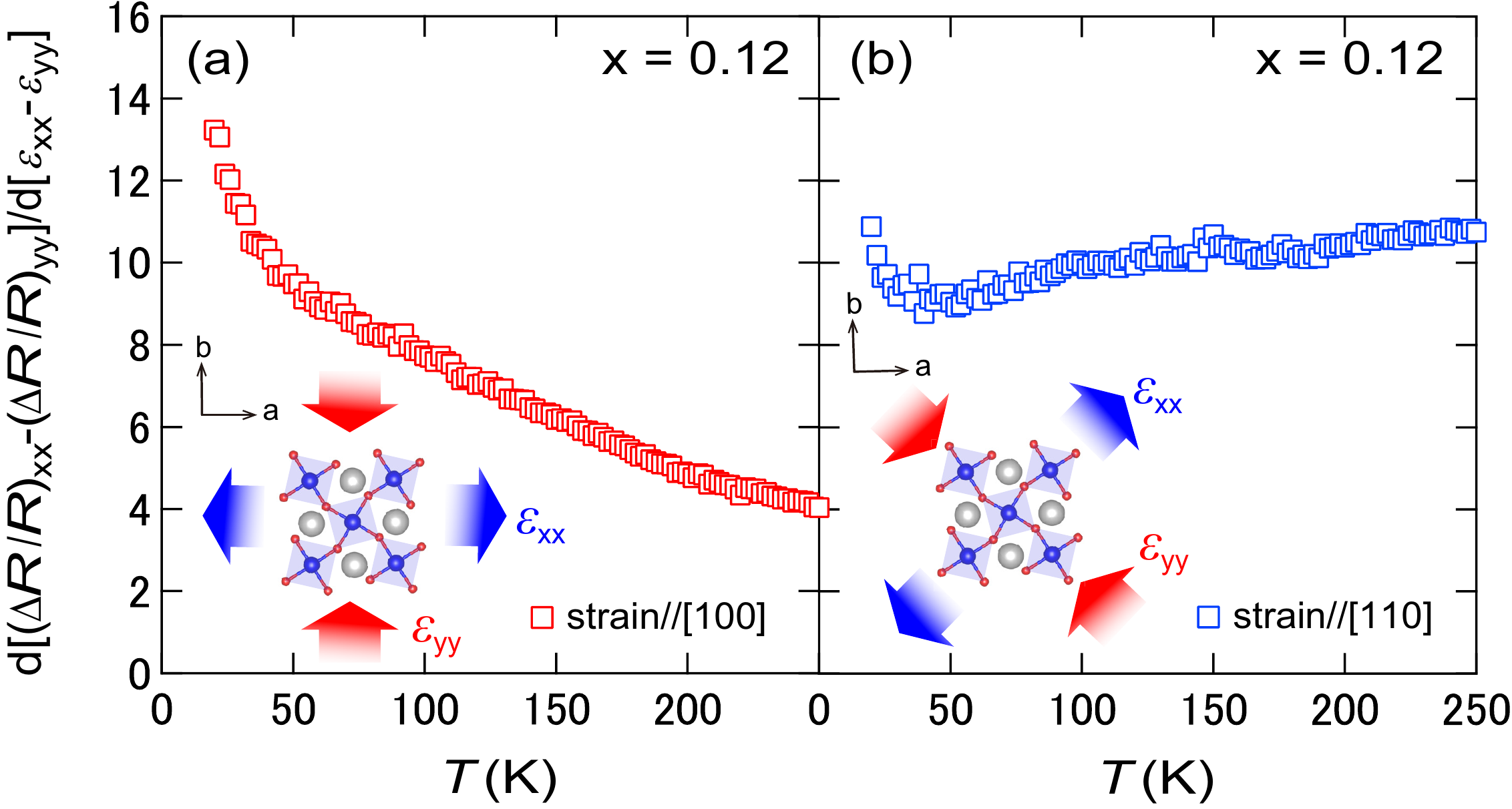}
	\caption{Temperature dependence of nematic susceptibility in Sr$_2$Ir$_{0.88}$Rh$_{0.12}$O$_4$. (a) Strain is applied along [100] direction.  (b) Strain is applied along [110] direction. The insets illustrate the directions of applied strain and the IrO$_6$ octahedrons.}
	\label{fig:elasto}
\end{figure}

The results of elastoresistance measurements along two different strain directions are shown in Figs.\,\ref{fig:strain}(a, b) and (c, d). The linearity of the resistance change with respect to the strain observed for both $x$ and $y$ directions indicates that the anisotropy $\eta$ is proportional to the strain.  The doping level of the samples is $x=0.12$, where the long-range antiferromagnetic and hidden-order transition temperatures are expected to be $\sim 30$\,K and $\sim150$\,K (see Fig.\,3). As shown in Fig.\,\ref{fig:elasto}(a), the temperature dependence of the nematic susceptibility $\chi_{\rm nem} = d(\Delta R_{xx}/R_{xx}-\Delta R_{yy}/R_{yy})/d(\varepsilon_{xx}-\varepsilon_{yy})$ along [100] show a gradual increase with decreasing temperature, but it does not show any divergent behavior near the putative hidden-order transition around 150\,K. The data for [110] direction [Fig.\,\ref{fig:elasto}(b)], which is the along the nematic director expected from the torque measurements, show even weaker temperature dependence with no discernible anomaly around 150\,K. In both directions, an enhancement is observed below 30\,K below the N\'{e}el temperature $T_N^L$, where the ground state is insulating. The observed absence of the divergent behavior in two directions, which correspond to two irreducible representations $B_{1g}$ and $B_{2g}$ in the tetragonal $D_{4h}$ system, provides evidence against even-parity nematic order in the hidden-order phase of Sr$_2$Ir$_{1-x}$Rh$_x$O$_4$. Therefore, this implies that the rotational symmetry breaking state revealed by the magnetic torque is of odd parity, as will be discussed later.

\section{Discussion}
 
Our main experimental findings for pure and  Rh-substituted Sr$_2$IrO$_4$ are threefold:  First, the in-plane torque magnetometry provides thermodynamic evidence of the nematic transition that lowers the rotational symmetry of the system from $C_4$ to $C_2$ at $\sim T_{\Omega}$. Second, the growth of $B_{2g}$ two-fold oscillations indicate a nematic director along the Ir-O-Ir [110] bond direction, in contrast to previous proposals~\cite{Zhao16,Jeong17}. Third, the elastoresistance shows no diverging nematic susceptibility towards $T_{\Omega}$. We point out that these results are consistent with the anapole state, which has been proposed by SHG and polarized neutron measurements~\cite{Zhao16,Jeong17}. 
It should be noted that the anapole order parameter breaks inversion symmetry while magnetic field preserves it. Thus, the coupling between the anapole moment and external magnetic field is not linear but arises from a second-order effect. This coupling gives rise to two-fold magnetic anisotropy, which is detected as two-fold oscillation of the in-plane magnetic torque.  Moreover, below $T_{\Omega}$, the amplitude of the two-fold oscillations  increases nearly linearly.  This implies that the nematicity is not a primary order parameter but a secondary one of the hidden order state.

In Fe-based superconductors, including Ba(Fe$_{1-x}$Co$_x$)$_2$As$_2$, BaFe$_2$(As$_{1-x}$P$_x$)$_2$ and FeSe$_{1-x}$S$_x$,  the  nematic susceptibility exhibits diverging behavior toward the nematic transition temperature~\cite{Chu12,Kuo16,Hosoi16}.  In addition, the divergent nematic susceptibility has also been reported in a cuprate  (Bi,Pb)$_2$Sr$_2$CaCu$_2$O$_{8+\delta}$ near the pseudogap critical point~\cite{Ishida20b}.   Therefore the absence of the divergent nematic susceptibility with approaching $T_{\Omega}$ despite the presence of thermodynamic nematic transition is in sharp contrast to the nematic transition reported in other strongly correlated electron systems.   One may suspect that electrons couple with lattice very weakly in \SIR.  However, such a possibility is unlikely because the inherently strong spin-orbit interaction in \SIR~is expected to give rise to a strong coupling between the electrons and the lattice.    The present results of \SIR~ suggest that the hidden order state is not a simple ferroic nematic ordered phase characterized by the wave vector $\bm{q}=0$.   
 
We stress that the absence of divergent nematic susceptibility is consistent with the odd parity order because of the following reasons. In the case of even-parity ferroic nematic order, the elastoresistance measurements can sensitively detect the divergent behavior of nematic susceptibility toward the transition temperature. If the order is of odd parity, on the other hand, the divergence may not be observed from the elastoresistance, because the linear coupling between the strain and the odd-parity order parameter is prohibited.   In fact, in the Landau theory, coupling of anapole order and strain is described as 
\begin{align}
	F = F_{\rm lat} + F_{\rm ana} + \gamma_1 T_x T_y \varepsilon_{xy}
	+ \gamma_2 \left(T_x^2 - T_y^2 \right) \left(\varepsilon_{xx} - \varepsilon_{yy}\right), 
	\label{Landau}
\end{align}
where ${\bm T} = (T_x, T_y)$ is a polar vector order parameter for the anapole order in $E_u$ representation, $\varepsilon_{ij}$ are strain, $\gamma_1$ and $\gamma_2$ are coupling constants, $F_{\rm lat}$ and $F_{\rm ana}$ are free energy of the lattice strain and anapole order, respectively~\cite{Varma09}. 
Any linear coupling of anapole order parameter and strain, $T_i \varepsilon_{jk}$, is prohibited because $T_i$ and $\varepsilon_{jk}$ have different spacial inversion and time-reversal parity; anapole order is odd-parity while the strain is even-parity. 
The absence of the linear couplings reveals no divergent susceptibility when approaching the anapole transition $T_\Omega$ from high temperatures. This is in contrast to the even-parity nematic case whose order parameters couple linearly to the strain.  
On the other hand, bond-directional anapole order parameter grows below $T_{\Omega}$ as $|T_x| = |T_y| \propto (T_{\Omega}-T)^{1/2}$ in the mean field region. Through the coupling term in Landau free energy, nematicity $\varepsilon_{xy} \propto T_{\Omega} -T$ is induced below $T_{\Omega}$. These behaviors expected in the mean field region of anapole order are compatible with the results of magnetic torque and elastoresistance measurements.

We now discuss the direction of the anapole moment and loop current pattern in the hidden order state.   The polar toroidal moment that directs to the $a$ axis and the simplest  loop current pattern  illustrated in Fig.\,\ref{fig:anapole}(a) have been proposed in the previous SHG and polarized neutron diffraction studies~\cite{Zhao16,Jeong17}.  In our torque magnetometry, however, the two-fold oscillations below $T_\Omega$ follow the functional form $\tau_{2\phi}=A_{2\phi}\cos 2\phi$.  This clearly indicates  $\chi_{aa}=\chi_{bb}$ and $\chi_{ab} \neq 0$ in Eq.(1), i.e. the emergence of the nematicity whose anisotropy axis is along the Ir-O-Ir direction, which is referred as bond nematicity.   The direction of the nematic director is the same as that of anapole moment.  Therefore,  the torque magnetometry reveals that anapole moment is directed to [110] direction, 45$^{\circ}$ rotated from the $a$ axis, as illustrated  in Fig.\,\ref{fig:anapole}(b). 

Among the resemblances of crystallographic, magnetic and  electronic structures between Sr$_2$IrO$_4$ and La$_2$CuO$_4$, the formation of pseudogap with the Fermi arc in electron-doped Sr$_2$IrO$_4$ has drawn a great deal of attention, because the pseudogap is one of the most prominent, and most discussed features of the cuprates.    Although the pseudogap is not observed in hole-doped Sr$_2$IrO$_4$ in the present $x$ range~\cite{Cao16, Louat18},  the emergence of rotational symmetry breaking is a remarkable common feature of  two different Mott insulating systems.   Thus, detailed comparisons of the nematicity between the cuprates and the iridate could reveal important information on the peculiar electronic properties of both systems.   

It has been reported that the nematic director in YBa$_2$Cu$_3$O$_{6+\delta}$  and (Bi,Pb)$_2$Sr$_2$CaCu$_2$O$_{8+\delta}$ orients along the Cu-O-Cu bond direction~\cite{Sato17,Auvray19}, while it orients along the diagonal direction of the CuO$_2$ square lattice in HgBa$_2$CuO$_{4+\delta}$~\cite{Murayama19}.  Although the difference of the nematic direction among cuprates remains an open question,  it may be due to the number of CuO$_2$ planes in the unit cell~\cite{Mangin-Thro17}.  Therefore, the comparison between the monolayer cuprate HgBa$_2$CuO$_{4+\delta}$ and the single layered iridate Sr$_2$IrO$_4$ is more relevant.   In HgBa$_2$CuO$_{4+\delta}$,  polarized neutron diffraction experiments have revealed an unusual ${\bm q}$ = 0 magnetic order  below the pseudogap temperature, which has been discussed in terms of loop current-like magnetism breaking both time-reversal and parity symmetries~\cite{Li08}. 
The in-plane torque magnetometry in  HgBa$_2$CuO$_{4+\delta}$ reported that the nematic director orients along the diagonal direction of the CuO$_2$ square lattice, similar to Fig.\,\ref{fig:anapole}(a).   Thus  the nematic director in the hidden order phase of  Sr$_2$IrO$_4$ is 45$^{\circ}$ rotated  from that in the pseudogap state of  HgBa$_2$CuO$_{4+\delta}$.  Therefore, if the loop current-like magnetism is an origin of the nematicity, the loop current pattern of Sr$_2$IrO$_4$, which is illustrated in Fig.\,\ref{fig:anapole}(b),  must be different from that of HgBa$_2$CuO$_{4+\delta}$.  While  the intra-unit cell loop-current flows along the Cu-O-Cu direction in the CuO$_2$ square in HgBa$_2$CuO$_{4+\delta}$, it flows only in one of the diagonal directions in the IrO$_4$ square in both pure and Rh-doped Sr$_2$IrO$_4$~\cite{Chatterjee17b,
Scheurer18, Sachdev19}.

The interpretation of the two-fold oscillations in the AFM ordered states is rather complicated.  The rapid suppression of the amplitude of two-fold oscillations with $B_{2g}$ symmetry below $T_N^S$ may be due to the segmentation of the domain structure by the short range magnetic order. Further detailed study is required to clarify how the short range magnetic order influences the anapole moment.  As shown in Figs.\,\ref{fig:amplitude}(a)-(c), two-fold oscillations with $B_{1g}$ symmetry appear below the long range AFM ordering temperatures.  This may be due to the coupling between pseudospin and lattice~\cite{Porras19}.

\bigskip
\section{Conclusions}
In summary, we measured the in-plane magnetic torque and elastoresistance in a spin-orbit coupled Mott insulator \SIR.  The measurements of in-plane anisotropy of the magnetic susceptibility reveal the emergence of the nematic phase with broken rotation symmetry distinctly above the N\'{e}el transition. On the other hand, nematic susceptibility exhibits no diverging behavior with approaching the nematic critical point. These results provide bulk evidence for the odd-parity hidden order state suggested by the SHG.  The results, along with the report of the polarized neutron diffraction, suggest that the hidden order phase is in an anapole state, which is composed of polar (magnetic) toroidal dipole induced by persistent loop currents.  These results may bear resemblance to the pseudogap state of underdoped cuprates, but the nematic director of \SIR~ is 45$^{\circ}$ rotated from that of monolayer cuprate  HgBa$_2$CuO$_{4+\delta}$,  implying the difference of the loop current patterns between the two systems.  The remarkable common features of the nematicity and distinct differences of the nematic director may be key to understanding the symmetry breaking phenomena in the hidden order state of the iridate and the pseudogap state of the cuprates. 

\section*{ACKNOWLEDGMENTS}
We thank T. Kimura, H. Kontani and E.-G. Moon for discussions. 
This work is supported by Grants-in-Aid for Scientific Research (KAKENHI) (Nos. JP18H01177, JP18H01178, JP18H01180, JP18H05227, JP18K13492, JP19H00649, JP20H02600, JP20H05159 and JP20K21139) and on Innovative Areas ``Quantum Liquid Crystals" (No. JP19H05824) from Japan Society for the Promotion of Science (JSPS), and JST CREST (JPMJCR19T5). GC acknowledges NSF support via grant DMR 1903888.

\end{document}